\documentstyle[11pt,paspconf,epsf]{article}
\def\edcomment#1{\iffalse\marginpar{\raggedright\sl#1\/}\else\relax\fi}
\reversemarginpar

\begin{document}

\def\ga{\mathrel{\raise.3ex\hbox{$>$\kern-.75em\lower1ex\hbox{$\sim$}}}}
\def\la{\mathrel{\raise.3ex\hbox{$<$\kern-.75em\lower1ex\hbox{$\sim$}}}}
\def\he#1{\hbox{${}^{#1}$He}}
\def\li#1{\hbox{${}^{#1}$Li}}
\def\be#1{\hbox{${}^{#1}$Be}}
\def\b#1#2{\hbox{${}^{#1#2}$B}}
\def\beq{\begin{equation}}
\def\eeq{\end{equation}}

\title{LiBeB and Big Bang Nucleosynthesis}

\author{Keith A. Olive }
\affil{Theoretical Physics Institute, School of Physics and Astronomy, 
University of Minnesota,
    Minneapolis MN 55455, USA}

\author{Brian D. Fields }
\affil{Department of Astronomy, University of Illinois,
Urbana IL 61801, USA}


\vskip -3in
\rightline{UMN-TH-1747/99}
\rightline{TPI-MINN-99/11}
\rightline{astro-ph/9902297}
\rightline{February 1999}
\vskip 2.3in

\begin{abstract}
The dual origin of population II \li7, in both big bang nucleosynthesis and
galactic cosmic-ray nucleosynthesis is discussed.  It is argued that with
additional \li6 data, stringent limits on the degree of \li7 depletion can be
obtained. \li7 depletion is also constrained by the concordance of big
bang predictions with observational
determinations of light element abundances. Stringent limits can also be
obtained for a fixed primordial D/H abundance. 
\end{abstract}

\section{Introduction}

The key link between galactic cosmic-ray and big bang nucleosynthesis (GCRN
and BBN) is the production of \li7.  Although the big bang produces each
of the element isotopes \li6, \li7, \be9, \b10, and \b11, their abundances
relative to GCRN production are nearly negligible, with the exception of 
\li7 for which BBN is the main source for the abundances determined from the
observation of population II halo stars.  In contrast, the other LiBeB
elements are all formed predominantly in GCRN.  Therefore, not only is
\li7 a common link relating the two process, but the interpretation of
the abundance of \li7 in halo stars may require an understanding of
both mechanisms.

One of the goals of this paper is to differentiate the sources of \li7 and in
particular the {\it primordial} abundance of \li7. Among the criteria that will
be used is the concordance of the primordial \li7 abundance with the other
light element data (\he4 and D), as well as the consistency with the galactic
evolution of \li6 and BeB. BBN produces significant amounts of 
D, \he3, \he4, and \li7 (see e.g. Olive 1999). Furthermore, in the standard BBN
model there is really only one free (undetermined) parameter, the
baryon-to-photon ratio,
$\eta$. Therefore, we can impose as a constraint the concordance between
the predicted abundances of the light elements with the observationally
determined abundances (with the exception of \he3 which is highly
dependent on the uncertainties of stellar and galactic evolution).
\li7, however, is also produced in spallation and fusion processes
associated with cosmic rays. By modeling the evolution of the observed
abundances of Be and B, we can determine the associated \li7 production 
(in a given model).  As we will see, \li6 has the potential to play a key
role in this type of investigation, since its GCRN production is very
similar to that of \li7 and is much less model dependent. However, the
current paucity of data (though we should celebrate the fact there is
some data) makes it difficult at this time to draw hard conclusions.

In what follows, we will very briefly review the key essentials of BBN
relevant to our discussion here. We will then discuss the relevant LiBeB data
and the dual origin for \li7. The main focus of the paper will be devoted to
the question of \li7 depletion and the limits that can be placed on depletion
from BBN and GCRN.

\section{Big Bang Nucleosynthesis}

As noted above, the standard BBN model is in fact a one-parameter model.
The key uncertainty in the prediction of the light element abundances is
the the baryon-to-photon ratio, $\eta$.  The other parameters normally
associated with BBN are now to a large extent fixed.  Experiments at LEP
have fixed the number of light neutrino flavors, $N_\nu = 3$, and the
neutron mean-life is now well measured to be $\tau_n = 886.7 \pm 1.9$ s (see
e.g. the Review of Particle Properties, 1998). The predicted abundances of 
the light elements are shown in Figure 1, which concentrates on
the range in
$\eta_{10}$ between 1 and 10 ($\eta_{10} = 10^{10} \eta$).
  The curves for
the \he4 mass fraction,
$Y$, bracket the computed range based primarily on the uncertainty of the
neutron mean-life. Uncertainties in the produced \li7 
abundances have been adopted from the results in 
Hata et al. (1996). Uncertainties in D and
\he3 production are small on the scale of this figure. 
The  boxes correspond to the observed abundances with 2$\sigma$
statistical uncertainties and the dashed boxes include systematic
uncertainties.

As one can see, the predicted abundances of the light elements are
roughly 
\begin{eqnarray}
Y_p & \approx & 0.23 - 0.25  \nonumber \\
{\rm D/H} & \approx & 3.4 - 20 \times 10^{-5}  \nonumber \\
\he3/{\rm H} & \approx & 1 - 2 \times 10^{-5}  \nonumber \\
\li7/{\rm H} & \approx & 1 -2 \times 10^{-10}   
\end{eqnarray}
To test the concordance of the BBN predictions with the observations
it has become useful to use a maximum likelihood method as described in
detail in Fields et al. (1996).  Shown in Figure 2a, are individual likelihood
distributions for \he4 and \li7. To obtain these distributions, the
observed values of $Y_p = 0.238 \pm 0.002 \pm 0.005$ (Olive, Skillman, \&
Steigman 1997, Fields \& Olive 1998) and
\li7/H $= (1.6
\pm 0.1) \times 10^{-10}$ (Molaro, Primas, \& Bonifacio 1995, Bonifacio \&
Molaro 1997) were adopted.
\li7 exhibits a two-peaked distribution due to the local minimum in the
predicted abundance versus
$\eta$.  These two elements are shown because of their relative lack of
model dependent galactic evolutionary effects, in contrast to D
and \he3.  In Figure 2b, the total likelihood function is shown (the product of
the two distributions).  As one can see there is broad agreement (based on
\he4 and
\li7) in the range
$1.55 < \eta_{10} < 4.45$.

\begin{figure}[t]
\epsfysize=4.7in
\epsfbox{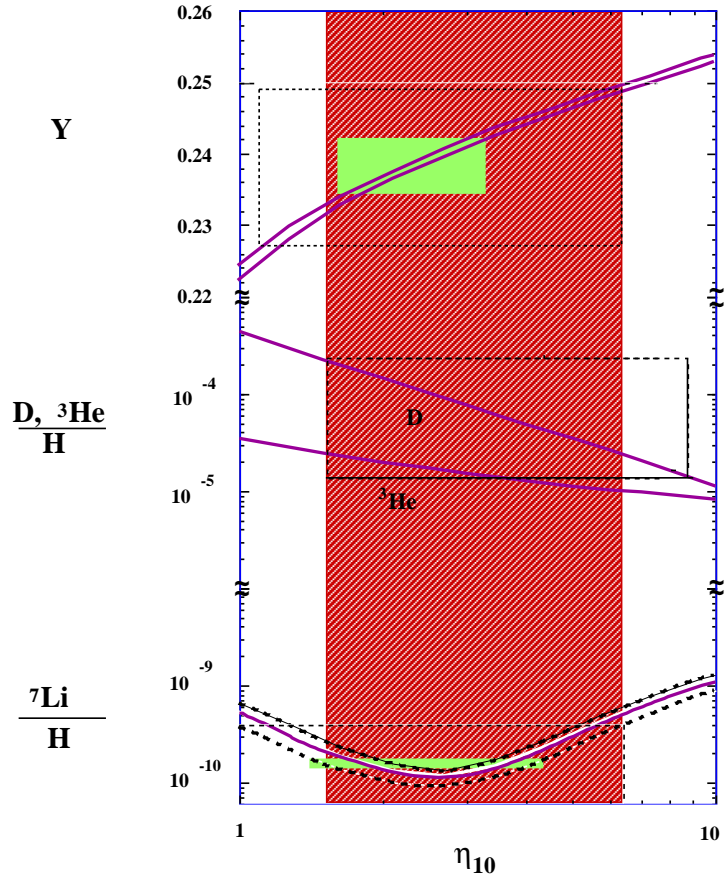} 
\caption{{The light element abundances from big bang
nucleosynthesis as a function of $\eta_{10} = 10^{10}\eta$.}}
\label{nuc8}
\end{figure}

In Figure 3a, the total likelihood distribution
including a high value of deuterium D/H = (2.0 $\pm 0.5) \times 10^{-5}$ is
shown (Songaila et al. 1994, Carswell et al. 1994, Webb et al. 1997, Tytler
et al. 1999). Now, the concordance is limited to a narrower range in $\eta$,
$1.5 < \eta_{10} < 3.4$ as can be seen from the total likelihood distribution
shown in Figure 3b. Similarly, in Figure 4, the distribution is shown for D/H
= 3.4 $\pm 0.3 \times 10^{-5}$ (Burles \& Tytler, 1998a,b). Notice the relative
scale in the two distributions (3b and 4b). The low value of D/H is compatible
with \he4 and \li7 at the 2 $\sigma$ level in the range $4.2 < \eta_{10} <
5.6$.

In contrast to the isotopes discussed above, big bang nucleosynthesis also
produces some \li6, Be, and B.  In the favored range of 1.5 -- 4.5 for
$\eta_{10}$, the big bang produces roughly (Thomas et al. 1993), 
\begin{eqnarray}
\li6/{\rm H} & \approx & 2 - 9 \times 10^{-14} \nonumber \\
\be9/{\rm H} & \approx & 0.04 - 2 \times 10^{-17}  \nonumber \\
\b10/{\rm H} & \approx & 0.5 - 3 \times 10^{-19}  \nonumber \\
\b11/{\rm H} & \approx & 0.02 - 1 \times 10^{-16}   
\end{eqnarray}
Because these abundances are far below the observed abundances found in Pop II
halo stars, (\li6/H $\approx$ few $\times ~10^{-12}$, \be9/H $\sim 1 - 10
\times 10^{-13}$, and B/H $\sim 1 - 10 \times 10^{-12}$), it is generally
recognized that these isotopes are not of primordial origin, but rather have
been produced in the Galaxy, through cosmic-ray nucleosynthesis.

\begin{figure}[h]
\plottwo{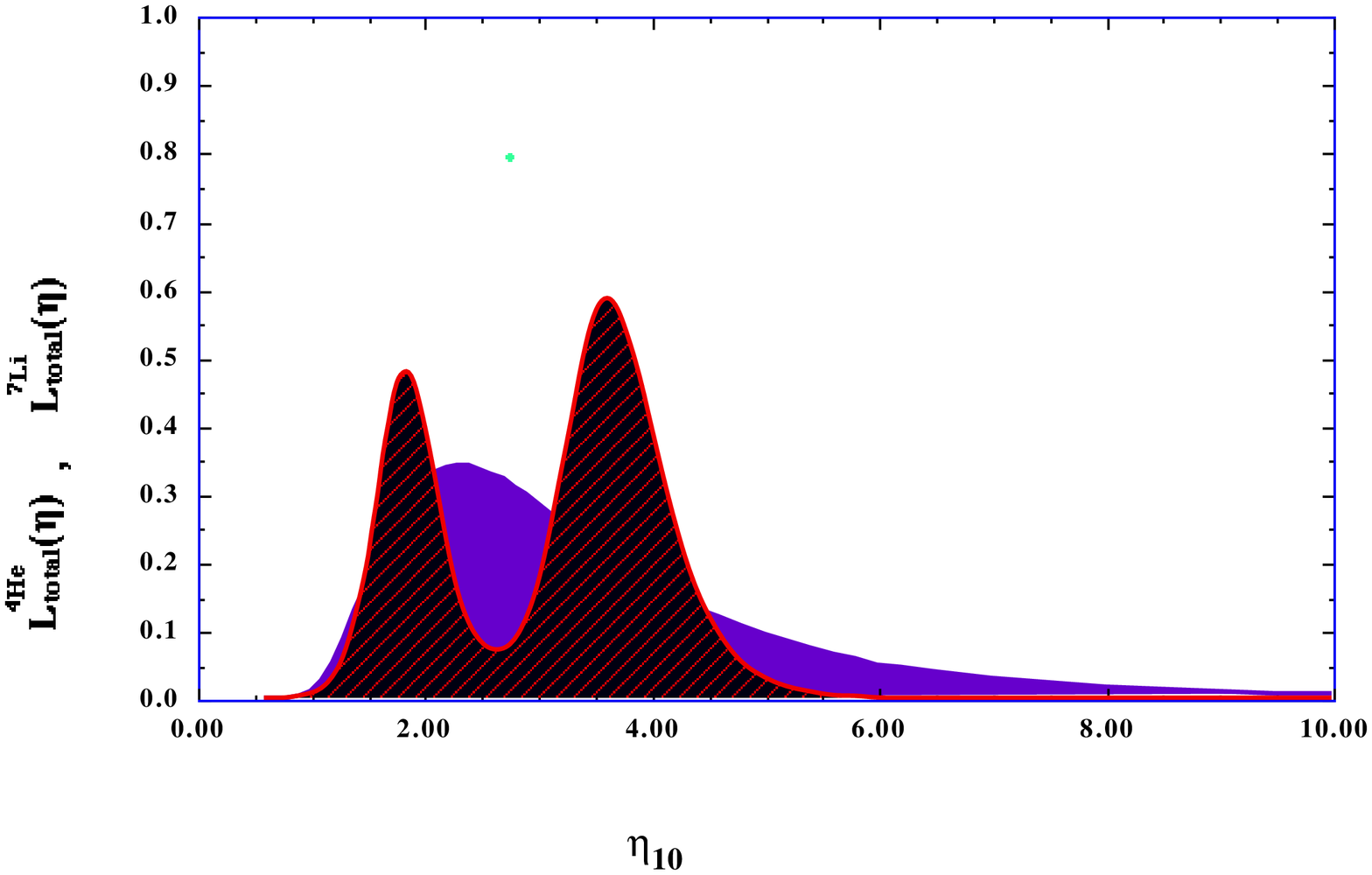}{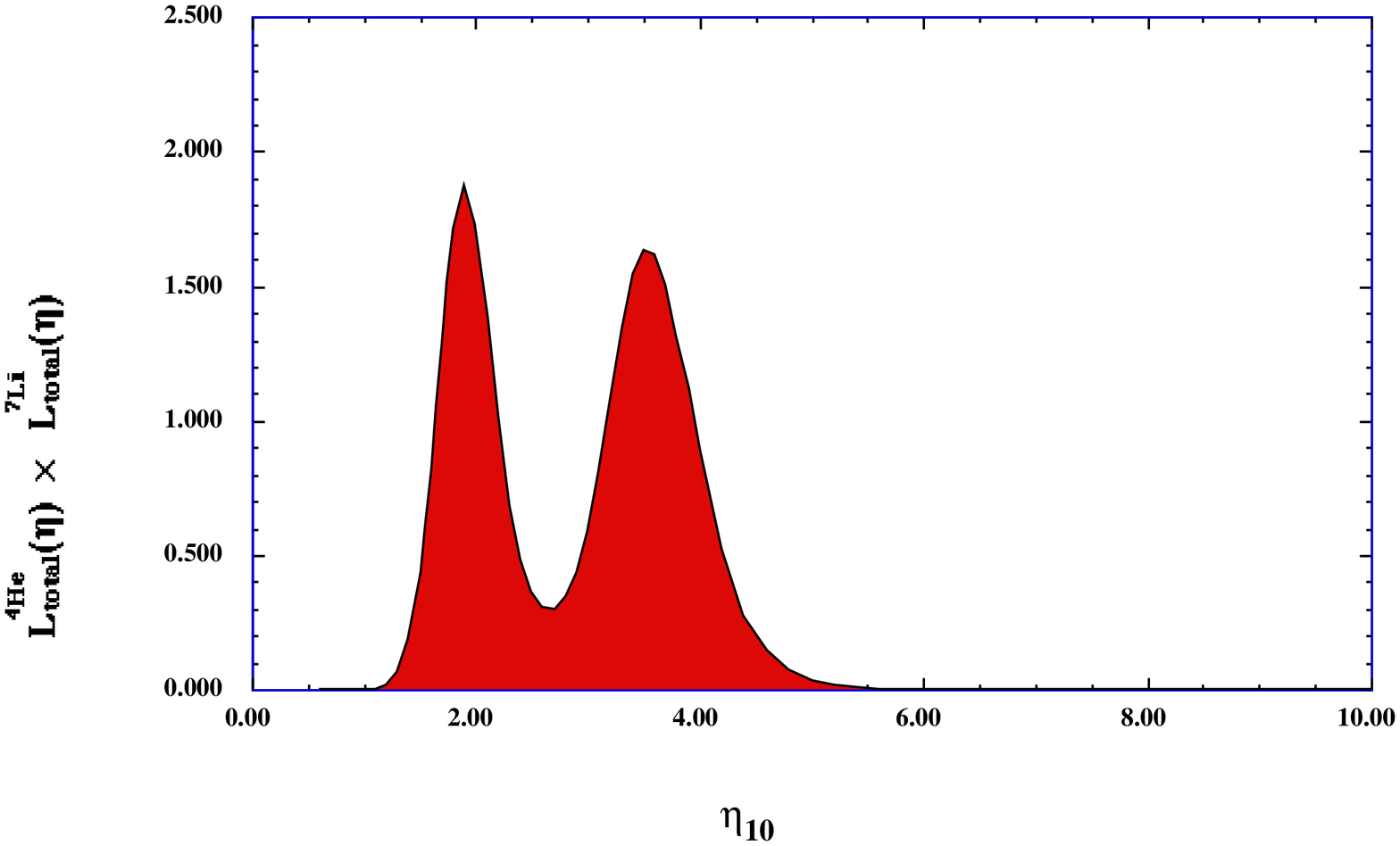} 
\caption{ (a) -  Likelihood distribution for each of \he4 and
\li7, shown as a  function of $\eta$.  The one-peak structure of the \he4
curve corresponds to the monotonic increase of $Y_p$ with $\eta$, while
the two peaks for \li7 arise from the minimum in the \li7 abundance
prediction; (b) - Combined likelihood for simultaneously fitting \he4 and \li7,
as a function of $\eta$.}
\end{figure}

\begin{figure}[h]
\plottwo{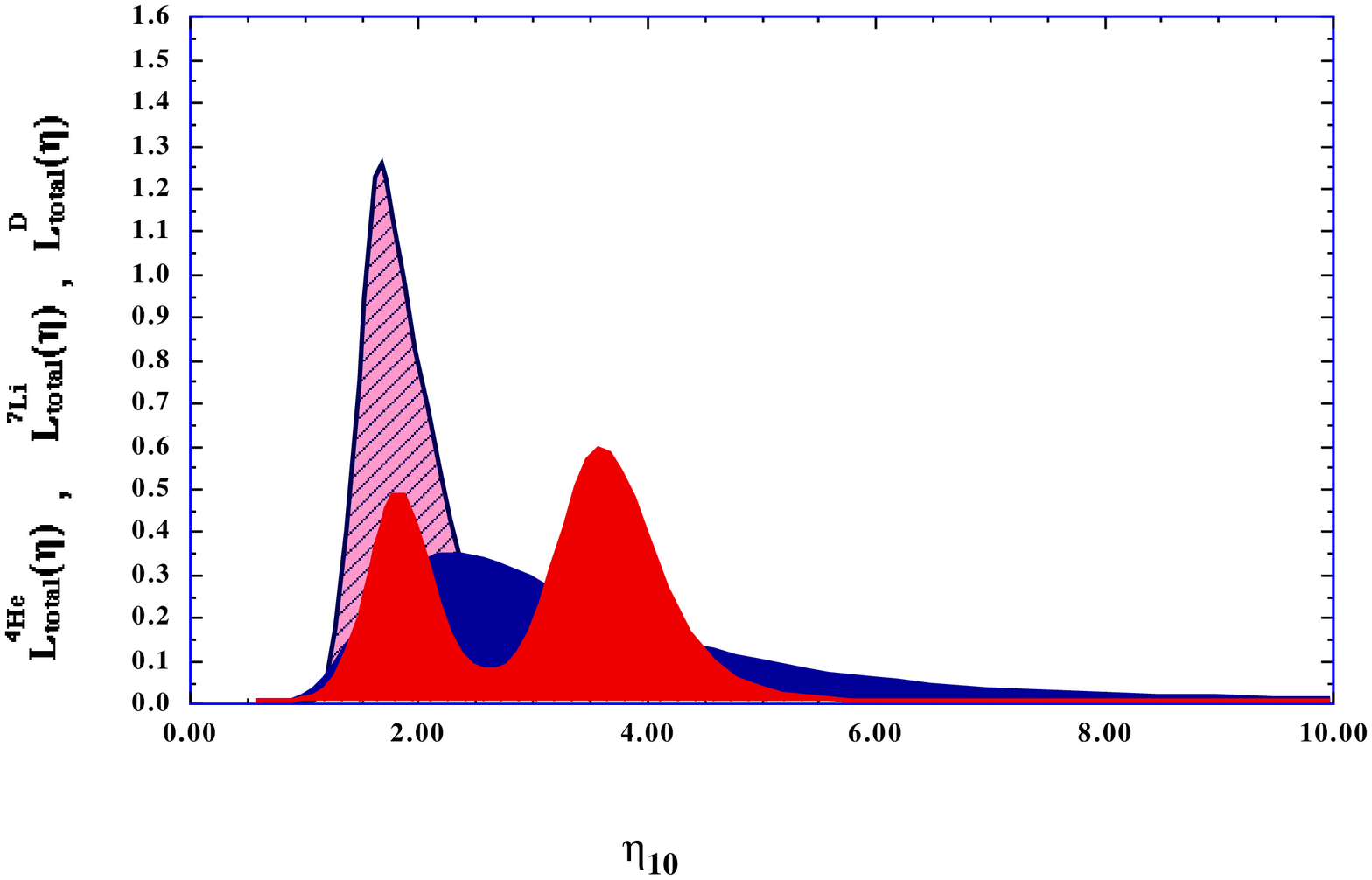}{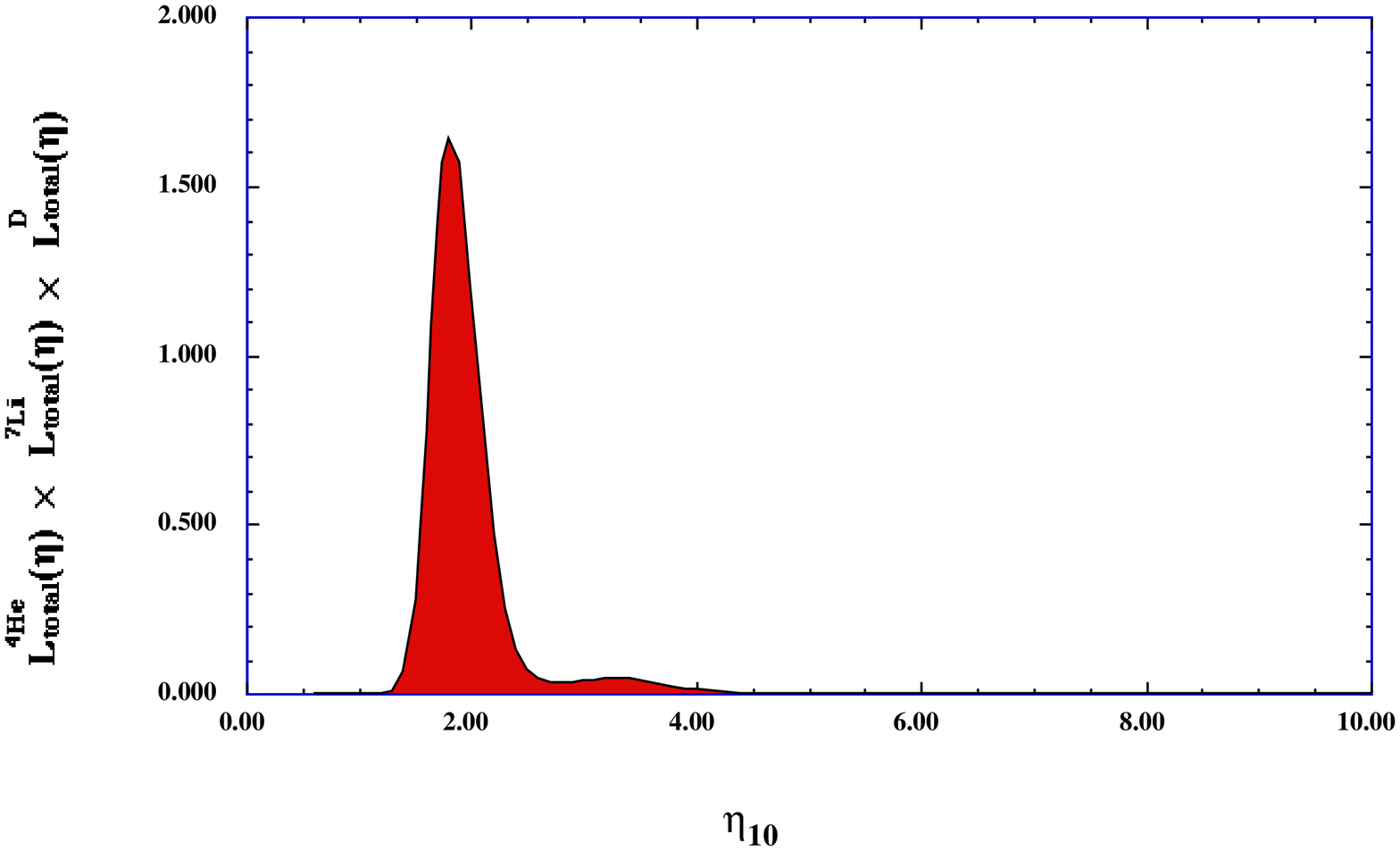} 
\caption{ (a) - As in Figure 2a, with the addition of the
likelihood  distribution for D/H assuming high D/H; (b) - The total
likelihood distribution with high D/H.}
\end{figure}

\begin{figure}[h]
\plottwo{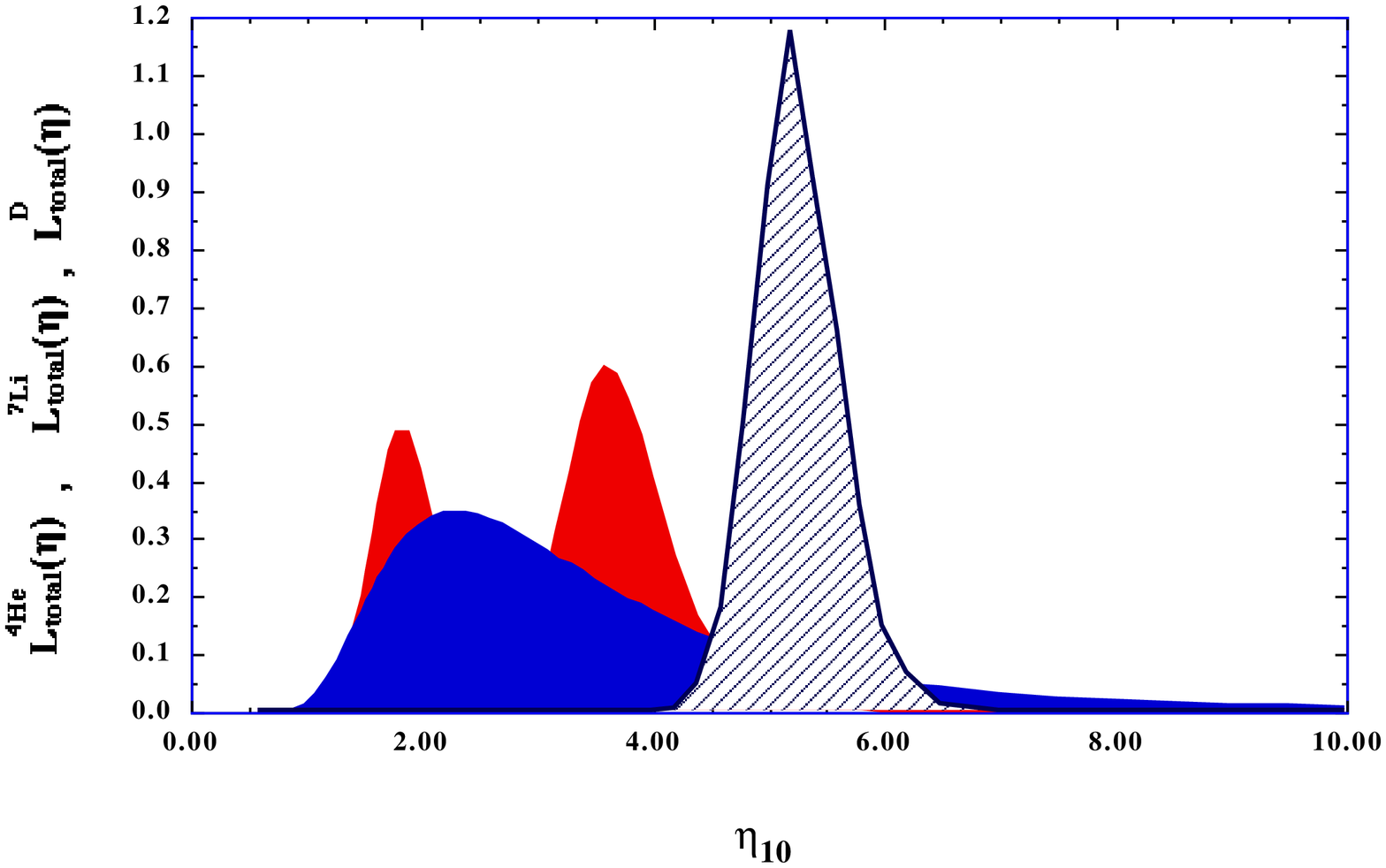}{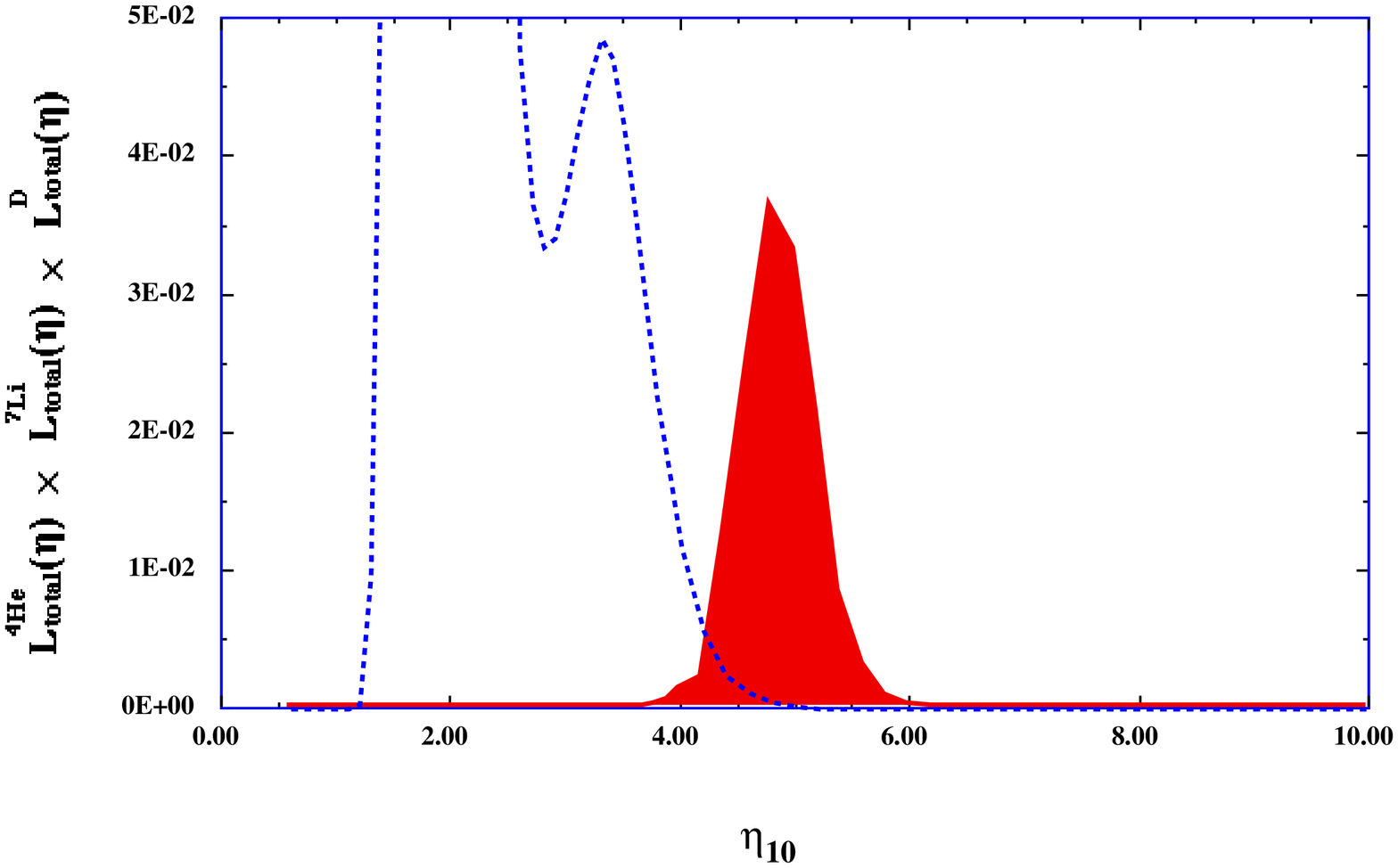} 
\caption{ (a) - As in Figure 2a, with the addition of the
likelihood  distribution for D/H assuming low D/H; (b) - The total
likelihood with low D/H. The dashed curve is the distribution
from Figure 3b.}
\end{figure}

\section{LiBeB Data}

The LiBeB data in population II halo stars is been discussed extensively in 
these proceedings and therefore, we can be brief here and concentrate on the
data as it concerns nucleosynthesis and the question of the primordial
abundance of \li7. 

 The \li7 data was reviewed by Molaro, and based on
the work of Molaro, Primas
\& Bonifacio (1995) and Bonifacio \& Molaro (1997) as well as more
recent results, he argued for a uniform abundance of \li7 with a value 
\beq
^7{\rm Li/H} \simeq (1.6 \pm 0.1) \times 10^{-10}
\eeq
absent of any trends with respect to either temperature or metallicity
([Fe/H]).

The \li6 abundance (reviewed here by Hobbs) has been improving over
time.  There are now two stars at low metallicity ([Fe/H] $\sim -2.3$) with
measured \li6/H (Smith, Lambert \& Nissen 1993,1998, Hobbs \& Thorburn
1994,1997, Cayrel et al. 1999):
\begin{eqnarray}
{\rm HD84937}: & &^6{\rm Li/Li_{\rm total}} = 0.054 \pm 0.011 \nonumber \\
{\rm BD26\deg3578}: & &^6{\rm Li/Li_{\rm total}} = 0.05 \pm 0.03
\end{eqnarray}
corresponding to \li6/H $\simeq 6 - 10 \times 10^{-12}$.
These values, together with the solar abundance (\li6 $\simeq 1.5 \times
10^{-10}$) lead to strong constraints on the evolution of \li6 and more
importantly on the nucleosynthesis of the LiBeB elements and the depletion of
\li7 as will be discussed below.

The BeB data was discussed by Duncan and Garci-Lopez.
The pop II BeB data is summarized by the tables below (Fields \& Olive 1999a).
The table below represents a compilation of data using a uniform set of stellar
parameters (effective temperature, surface gravity, and metallicity) to
extract abundances. Results for two methods are shown (see Fields \& Olive
1999a for more details). 

\begin{table}
\caption{Observed Pop II logarithmic slopes for Be versus Fe and O}
\label{tbl-1}
\begin{center}\scriptsize
\begin{tabular}{cccccc}
metal tracer & method & metallicity range & Be slope & B slope & B/Be
slope \\
\tableline
Fe/H & Balmer & $-3 \le [Fe/H] \le -1$ & $1.21 \pm 0.12$ & $0.65 \pm 0.11$
&     $-0.18 \pm 0.15$ \\
O/H & Balmer  & $-2.5 \le [O/H] \le -0.5$ & $1.76 \pm 0.28$ & $1.84 \pm
0.58$ & $-0.81 \pm 0.44$ \\
Fe/H & IRFM & $-3 \le [Fe/H] \le -1$ & $1.30 \pm 0.13$ & $0.77 \pm 0.13$
&     $0.01 \pm 0.14$ \\
O/H & IRFM  & $-2.5 \le [O/H] \le -0.5$ & $1.38 \pm 0.19$ & $1.35 \pm 0.30$
& $0.00 \pm 0.17$ \\
\end{tabular}
\end{center}
\end{table}

In standard GCRN (Meneguzzi, Audouze, \& Reeves 1971), \be9 is a secondary
isotope, and is expected to have a logarithmic slope of 2 with respect to
[Fe/H].  In the context of GCRN, B/H is also produced with a slope of 2,
but there is an additional source for boron from neutrino spallation in
supernovae (Hartmann 1999) which is primary (slope of 1). The [Fe/H] data shown
in the table, however, do not reflect the expectations of standard GCRN.  
As a result, many models have been developed over the last several years to
explain this data.  For a review of these, see Ramaty (1999).

The apparent failure of GCRN to account for the evolution of BeB vs. [Fe/H]
rests upon the assumption that at low metallicity, [O/Fe] is constant. 
However, new data (discussed here by Garci-Lopez) from Israelian et al. (1998)
indicates that in fact O/Fe is not constant.
If we take [O/Fe] =
$\omega_{\rm O/Fe}$[Fe/H], then we would expect up to an additive constant
(Fields \& Olive 1999a) 
\beq
[{\rm Be}]  = 2 (1 + \omega_{\rm O/Fe}) \ {\rm [Fe/H]}
\eeq 
Now for the Israelian et al.\ (1998) value of 
$\omega_{\rm O/Fe} = -0.31$, we would predict a Be slope which
is consistent with the data.  Galactic chemical evolution models with a
varying O/Fe were presented here by Fields.  As we will see below, these
models also affect the evolution of \li6 in a positive way (Fields \& Olive
1999b).

\section{Two-component \li7}

In order to test big bang nucleosynthesis, it is necessary to establish a
primordial abundance of \li7. The extraction of the primordial \li7 abundance
is complicated by two factors: the GCRN production of \li7 and the
depletion of \li7 in halo stars. If we for the moment ignore the depletion,
we must still ascertain what fraction of the observed \li7 is primordial.
In principle, we can use the abundance information on the other LiBeB isotopes
to determine the abundance of the associated GCRN produced \li7.
As it turns out, the boron data is problematic for this purpose, as there is
very likely an additional source for \b11, namely $\nu$-process
nucleosynthesis in supernovae as indicated above. In Walker et al. (1993), and
Olive \& Schramm (1992), the Be data was used to set a rough upper bound of
20-30\% to the fraction of GCRN produced \li7 on a star by star basis. 

There are in fact many stars for which both \li7 and \be9 have been detected.
Using this subset of the data, one can extract the primordial abundance of
\li7 in the context of a given model of GCRN.
For example, a specific GCRN model, predicts the ratio of Li/Be as a function
of [Fe/H].  Under the (plausible) assumption that all of the observed Be is
GCRN produced, the Li/Be ratio would yield the GCRN produced \li7 and could
then be subtracted from each star to give a set of primordial \li7 abundances.
This was done in Olive \& Schramm (1992) where it was found that the plateau
was indeed lowered by approximately 0.07 dex.
However, it should be noted that this procedure is extremely model dependent.
The predicted Li/Be ratio in GCRN models was studied extensively in Fields,
Olive \& Schramm (1994). It was found that Li/Be can vary between 10 and
$\sim 300$ depending on the details of the cosmic-ray sources and
propagation--e.g., source spectra shapes, escape pathlength
magnitude and energy dependence, and kinematics.

In contrast, the \li7/\li6 ratio is much better determined and far less model
dependent since both are predominantly produced by $\alpha-\alpha$ fusion
rather than by spallation. The obvious problem however, is the paucity of \li6
data.  As more \li6 data becomes available, it should be possible to obtain a
better understanding of the relative contribution to \li7 from BBN and GCRN.

\section{Li Depletion}

Stellar evolution models
have predicted depletion factors which differ widely, ranging
from essentially no depletion in standard models (for stars with $T \ga
5500$ K) to a large depletion (Deliyannis et al. 1990, 
Charbonnel et al. 1992). Depletion occurs when the base of the
convection zone sinks down and is exposed to high temperatures, $\sim 2
\times 10^6$ K for \li7 and $\sim 1.65 \times 10^6$ K for \li6 (Brown \&
Schramm 1988). In standard stellar models, the depletion of
\li7 is always accompanied by the depletion of \li6, though the
converse is not necessarily true. Below, the consequences of \li7 depletion
will be examined from both its effect on the galactic evolution of \li6 and
its effect on the concordance of BBN and the observations of D and \he4. 

\subsection{\li6 and depletion}

Model results (Fields \& Olive 1999b) for \li6 vs. Fe appear in 
Figure 5, for an
[O/Fe]-[Fe/H] Pop II slope 
$\omega_{\rm O/Fe} = -0.31$ (as discussed above)
and $\omega_{\rm O/Fe} = 0$ for comparison.
We see that GCRN does quite
well in reproducing both solar and Pop II \li6
when O/Fe is allowed to evolve in Pop II.
On the other hand, if O/Fe is constant, 
then the \li6-Fe slope is steeper and 
the model underproduces the Pop II \li6.
This is consistent with what is obtained for the evolution of BeB (Fields \&
Olive 1999a,c). 

\begin{figure}[h]
{\centering \leavevmode
\epsfxsize=5in \epsfbox{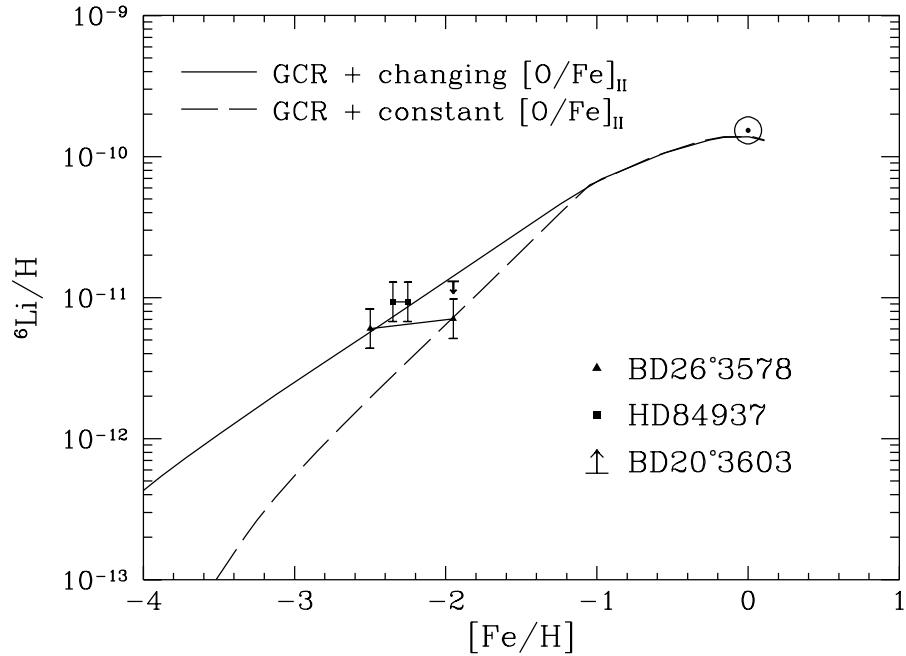}}
\caption{The \li6 evolution as a function of [Fe/H].
{\it Solid line}:  the ``revised standard'' GCRN model.
Here Fe is
scaled from the calculated O to fit the observed [O/Fe]--[Fe/H] slope.
{\it Dashed line}:  the GCRN model with 
Fe $\propto$ O in Pop II. The error bars on the points are 2 sigma
errors, and the spread in the points connected by lines show the
uncertainty due to stellar parameter choices.}
\end{figure}
It is also useful to compare the evolution of \li6 with that of Be or B.
The problem  again is the paucity of data. For HD 84937, data is available for
both \li6 and Be.   By comparing the solar ratio of
$\li6/{\rm Be}_\odot = 5.9$ and the value of this
ratio for HD84937, $\li6/{\rm Be} \simeq 80 $,
it appears that GCRN is consistent (though the observational uncertainties
remain large).  This data are good enough however, to exclude the possibility
of a purely primary (linear) evolution for [Be/H]
vs. [Fe/H] coupled with the expected linear evolution of
\li6 as a primary element due to $\alpha-\alpha$ fusion.  In this case one
would expect the
\li6/Be ratio to be constant, which  is
clearly not the case.  

In standard GCRN (with constant O/Fe at low
metallicities), \be9 is a secondary isotope, and given the linearity of
[\li6], one should expect that
\li6/Be is inversely proportional to Fe/H (i.e., to have
a log slope of -1).  However, if we take $\omega_{\rm O/Fe} \ne 0$, then
we would expect the Be evolution to be governed by Eq. (5) (Fields \& Olive
1999a,b) and
\beq
[\li6]  =  (1 + \omega_{\rm O/Fe}) \ {\rm [Fe/H]}
\eeq
so that
\beq
[\li{6}/{\rm Be}] = - (1 + \omega_{\rm O/Fe}) \ {\rm [Fe/H]}
\eeq
The Israelian et al.\ (1998) value of 
$\omega_{\rm O/Fe} = -0.31$, implies a dependence which
is consistent with the data.

The above comparison of the data as shown in Figure 5 to the
models do not take into account any depletion of \li6. There is
still a great deal of uncertainty in the amount of depletion for both \li6 and
\li7 as well as the relative depletion factor, $D_6/D_7$
(Chaboyer 1994, Vauclair and Charbonnel 
1995, Deliyannis et al. 1996, Chaboyer 1998, Pinsonneault et al.
1992, Pinsonneault et al. 1998, Pinsonneault 1999, Vauclair 1999). 
The observed lithium abundance can be expressed as 
\beq
{\rm Li}_{\rm Obs} = D_7 ( {\rm \li7_{BB}} + {\rm \li7_{CR}}) + 
D_6 ( {\rm \li6_{BB}} + {\rm \li6_{CR}})
\label{li}
\eeq
where the $D_{6,7} < 1$ are the \li{6,7} depletion factors.
The two lithium components due to big bang and cosmic ray production,
discussed in the previous section, are shown explicitly here. 
Given enough \li6 Pop II data, one could use the observed \li6 evolution
(1) to infer $\li7_{\rm CR}$ and thus $\li7_{\rm BB}$,
and (2) to measure \li6/Be and thereby constrain in more detail the 
nature of early Galactic cosmic rays both which would lead to a better
understanding of Li depletion in general.

In standard stellar models, Brown \& Schramm (1988) have argued that $D_6
\sim D_7^\beta$ with $\beta \approx 60$.  Clearly in this case any
observable depletion of \li7 would amount to the total depletion of \li6. 
Hence the observation of \li7 in HD84937 has served as a basis to limit
the total amount of \li7 depletion (Steigman et al. 1993, Lemoine et al.
1997, Pinsonneault et al. 1998). There are however, many models based
on diffusion and/or rotation which call for the depletion of \li6 and \li7
even in hot stars. 
The weakest constraint comes from assuming that depletion
occurs entirely due to mixing, so the destruction of
the Li isotopes is the same despite the greater fragility of 
\li6.
Because \li6/\li7 $\sim 1$ in cosmic-ray
nucleosynthesis, the observation of \li6 does exclude any model with
extremely large \li6 depletion if one requires the preservation of the Spite
plateau for \li7 up to [Fe/H] = -1.3 (Pinsonneault et al. 1998, Smith et al.
1998). However, barring an alternative source for the production of \li6, the
data are in fact much more restrictive. At the 2$\sigma$ level, the model
used to produce the evolutionary curve in Figure 5, 
would only allow a
depletion of \li6 by 0.15 dex ($D_6 > 0.7$); since $D_7 \ge D_6$, this
is also a lower limit to $D_7$.  

Further constraints on $D_7$ become available if we adopt a
model which relates \li6 and \li7 depletion.  E.g., 
if we use $\log D_6 = -0.19 + 1.94 \log
D_7$ as discussed in Pinsonneault et al. (1998), the data in the context
of the given model would not allow for any depletion of \li7. However, given
the observation al uncertainties, together with the uncertainties in the
stellar parameters it is possible to account for some Li depletion.  For
example, using the Balmer line stellar parameters, Fields
\& Olive (1999a) found 
$\omega_{\rm O/Fe} = -0.46 \pm 0.15$.  Using the value of -0.46, it was
determined that at  2$\sigma$ (with respect to the \li6 data) that $\log D_6 >
-0.32$ and would still limit $\log D_7 > -0.07$. Even under what most would
assume is an extreme O/Fe dependence of $\omega_{\rm O/Fe} = -0.61$,  \li6
depletion is limited to by a factor of 3.5 and corresponds to an upper limit
on the depletion of \li7 by 0.2 dex.  This is compatible with the upper limit
in Lemioine et al. (1997) though the argument is substantially different.

It should be clear at this point, that improved ($\equiv$ more) data on
\li6 in halo stars can have a dramatic impact on our understanding of
cosmic-ray nucleosynthesis and the primordial abundance of \li7.
Coupled with improved data on the O/Fe ratio in these stars, we would be
able to critically examine these models on the basis of their predictions
of \li6 and \be9.

\subsection{Constraints from BBN}
It is also in principle possible to constrain the degree of \li7 depletion
from the concordance of the light elements produced in BBN and the
observations. If \li7 depletion were significant, then the comparison of BBN
predictions to the observed abundances should be based on a \li7 abundance
which is greater than that determined in pop II halo stars. As a result of the
local minimum in the BBN  produced \li7 abundance (at about $\eta_{10} \sim
3$) the two most likely values of $\eta$ from \li7 (the twin peaks in the
likelihood distribution of Figure 2a), are split farther apart as is shown in
Figure 6a, where the assumed \li7 abundance is \li7/H $= (4.1 \pm 0.1) \times
10^{-10}$. This corresponds to a depletion factor of 0.4 dex, which was
recently argued to be an upper limit to the \li7 depletion (Pinnsoneault et
al. 1998, Pinnsoneault 1999). As one can see, the previous excellent agreement
between \li7 and \he4 (seen in Figure 2a) is lost.  The lack of agreement is
seen more quantitatively by comparing the two likely distributions of Figure 2b
and 6b.

When one also considers deuterium, the concordance is further disturbed. 
For high D/H, where the agreement between D/H and \he4 is good, there is
barely any overlap with Li as can be seen in Figure 7a.  The total likelihood
function in this case, shown in Figure 7b, can be compared with that in Figure
3b, which shows the likelihood distribution with high D/H and no \li7
depletion.

Finally, one can consider the case of low D/H. Here the concordance was never
very good.  \li7 depletion causes the high-$\eta$ peak of the Li distribution
to move to the right (to higher values of $\eta$), and for a depletion factor
of 0.4 dex as is assumed in Figure 8, the peak of the \li7 distribution lies
at values of $\eta$ larger than that predicted by low D/H.
Clearly for a smaller \li7 depletion factor, good agreement between \li7 and
D/H can be achieved.  In this case, \he4 is still somewhat problematic.

\begin{figure}[h]
\plottwo{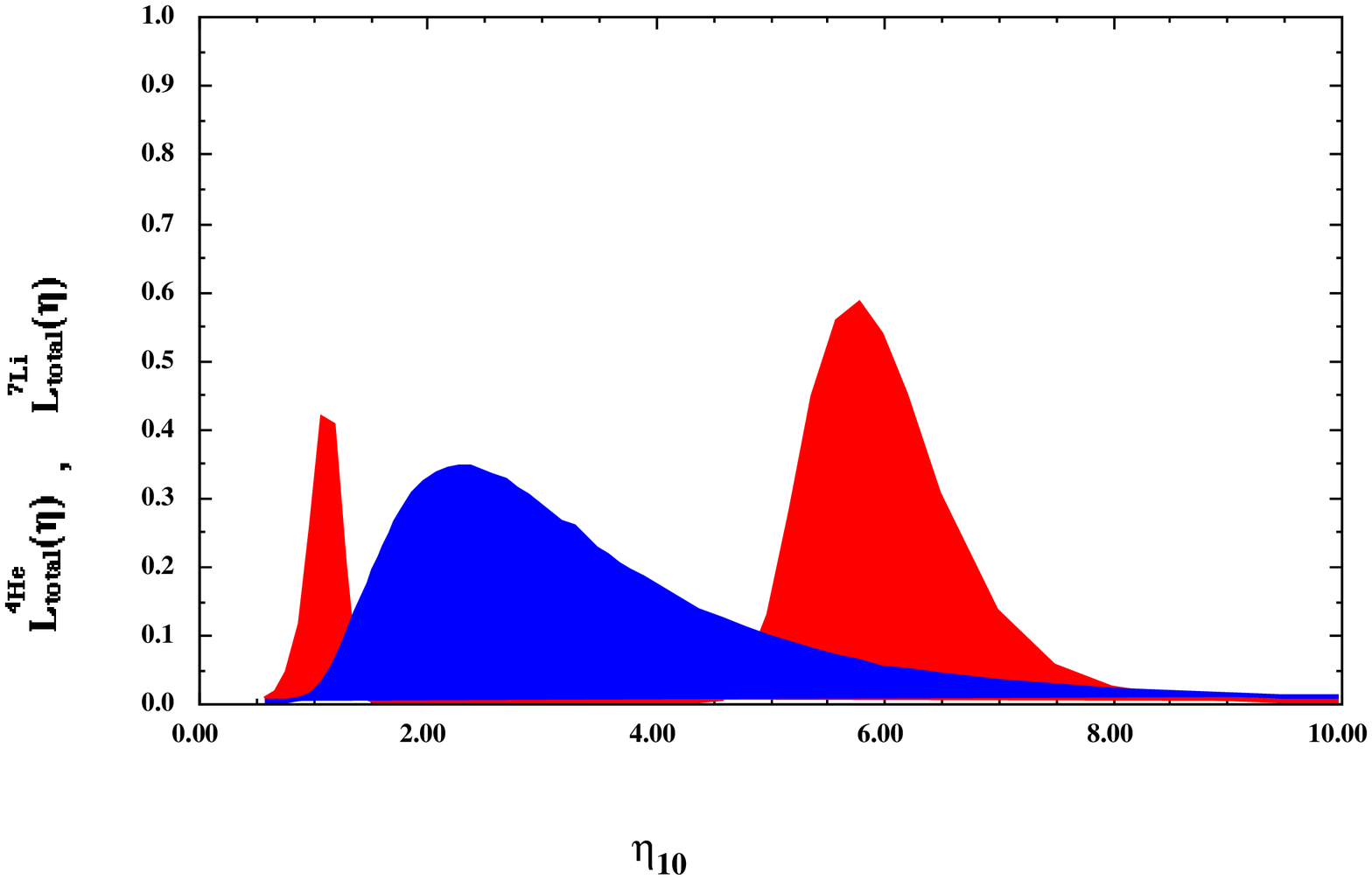}{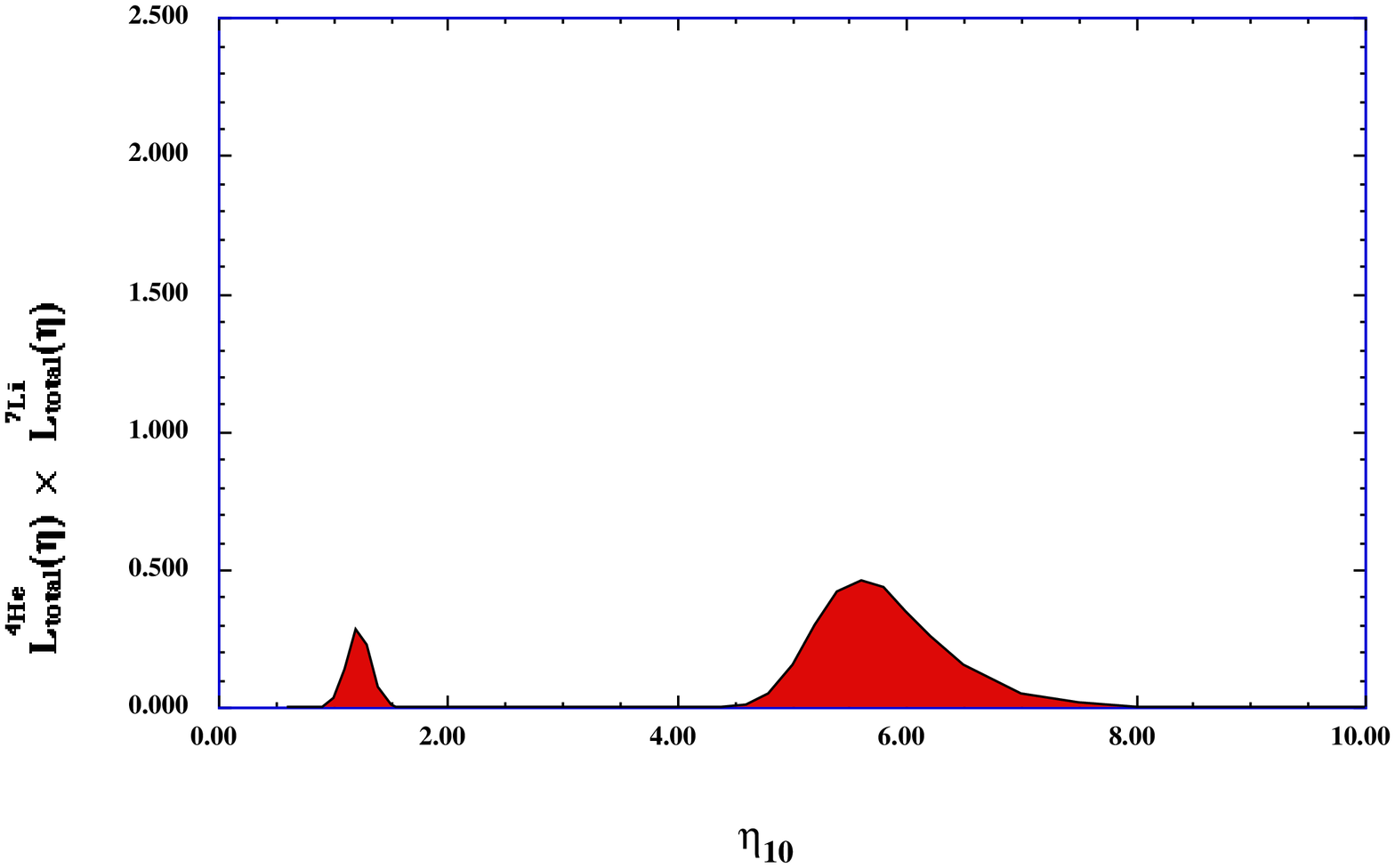} 
\caption{ (a) - As in Figure 2a, where the assumed primordial \li7 abundance
has been increased to take into account a possible 0.4 dex of depletion; (b) -
As in Figure 2b, with the increased primordial \li7.}
\end{figure}

\begin{figure}[h]
\plottwo{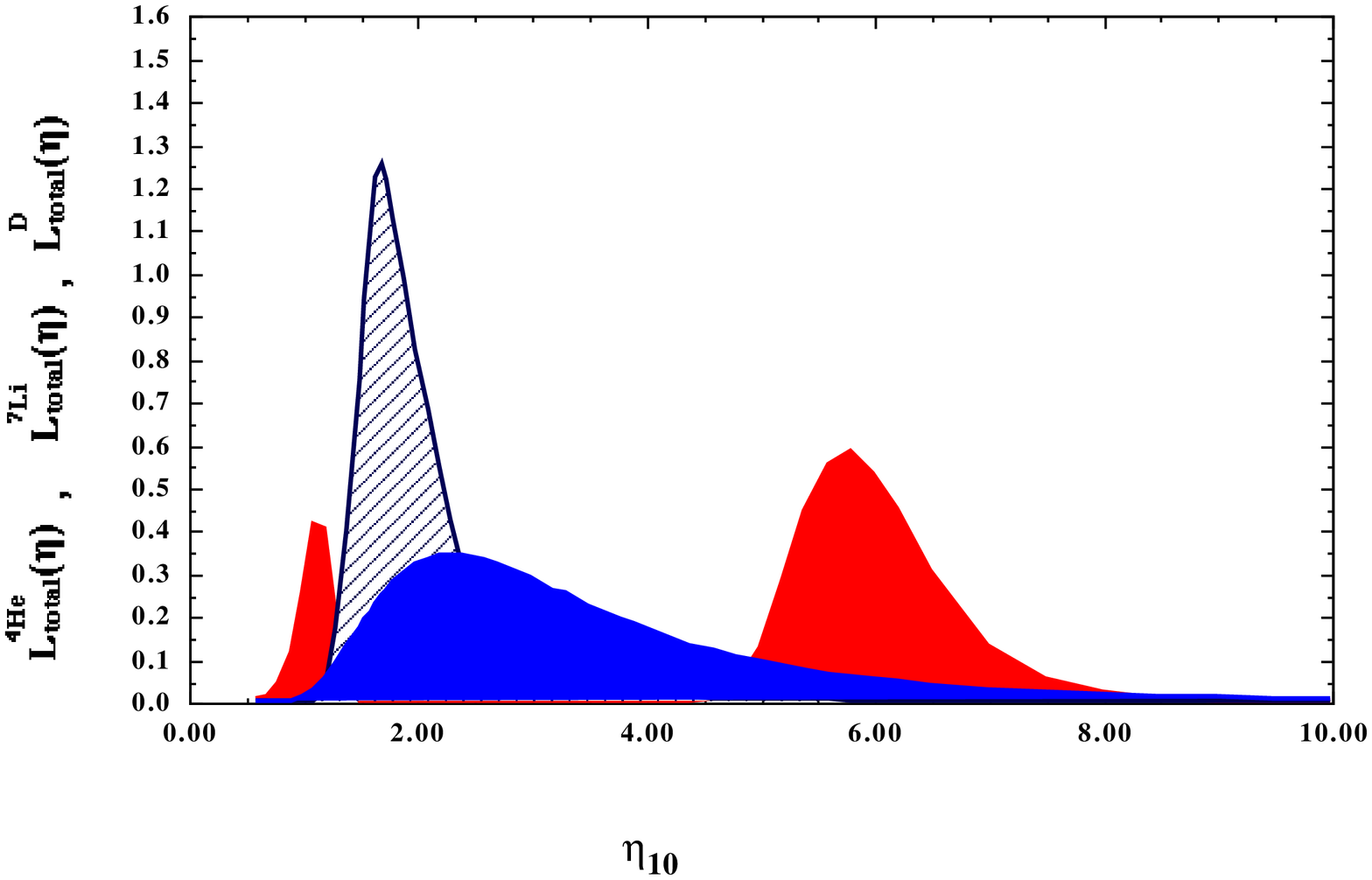}{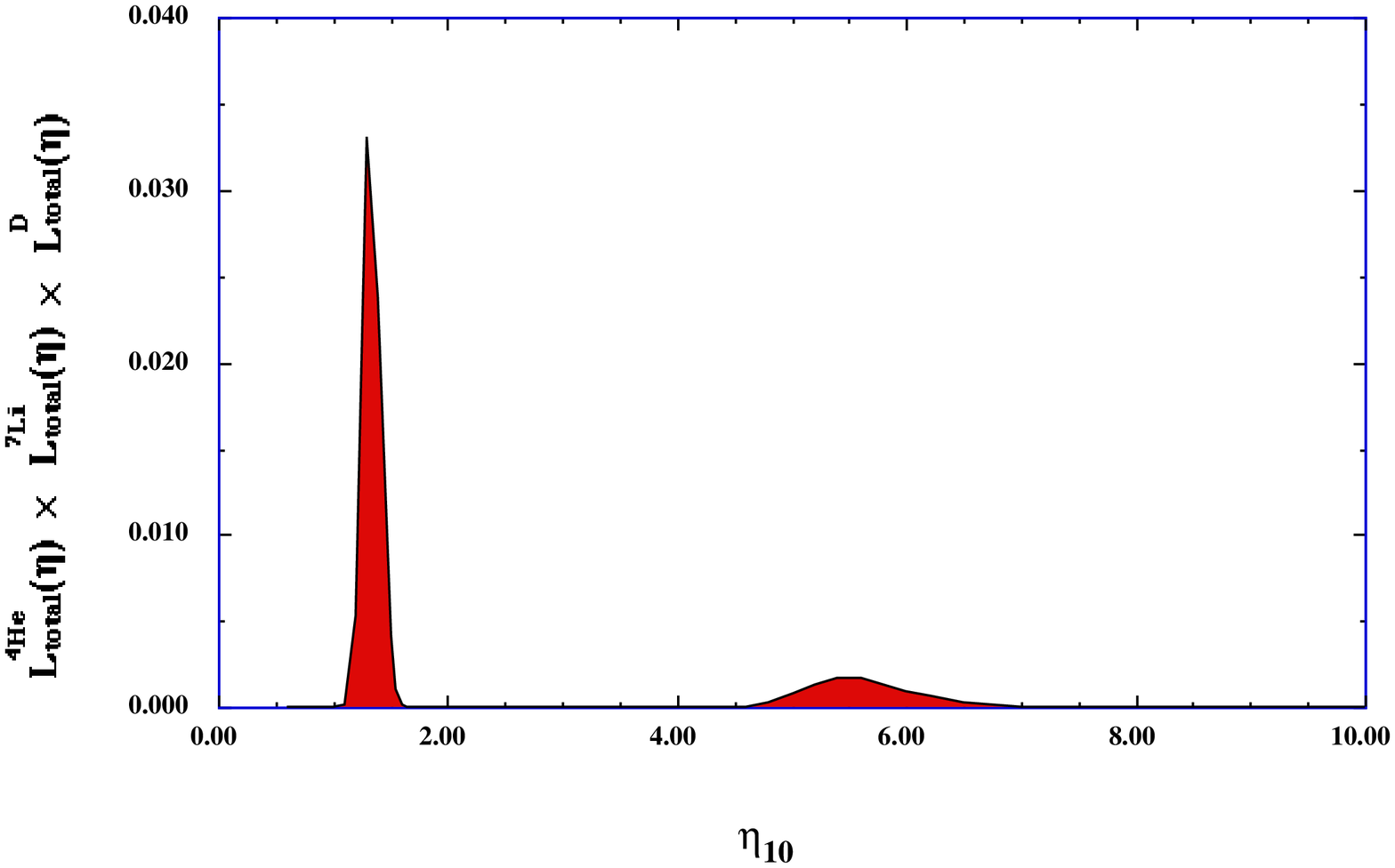} 
\caption{(a) - As in Figure 6a, with the addition of the
likelihood  distribution for D/H assuming high D/H; (b) - The total
likelihood distribution with high D/H.}
\end{figure}

\begin{figure}[h]
\plottwo{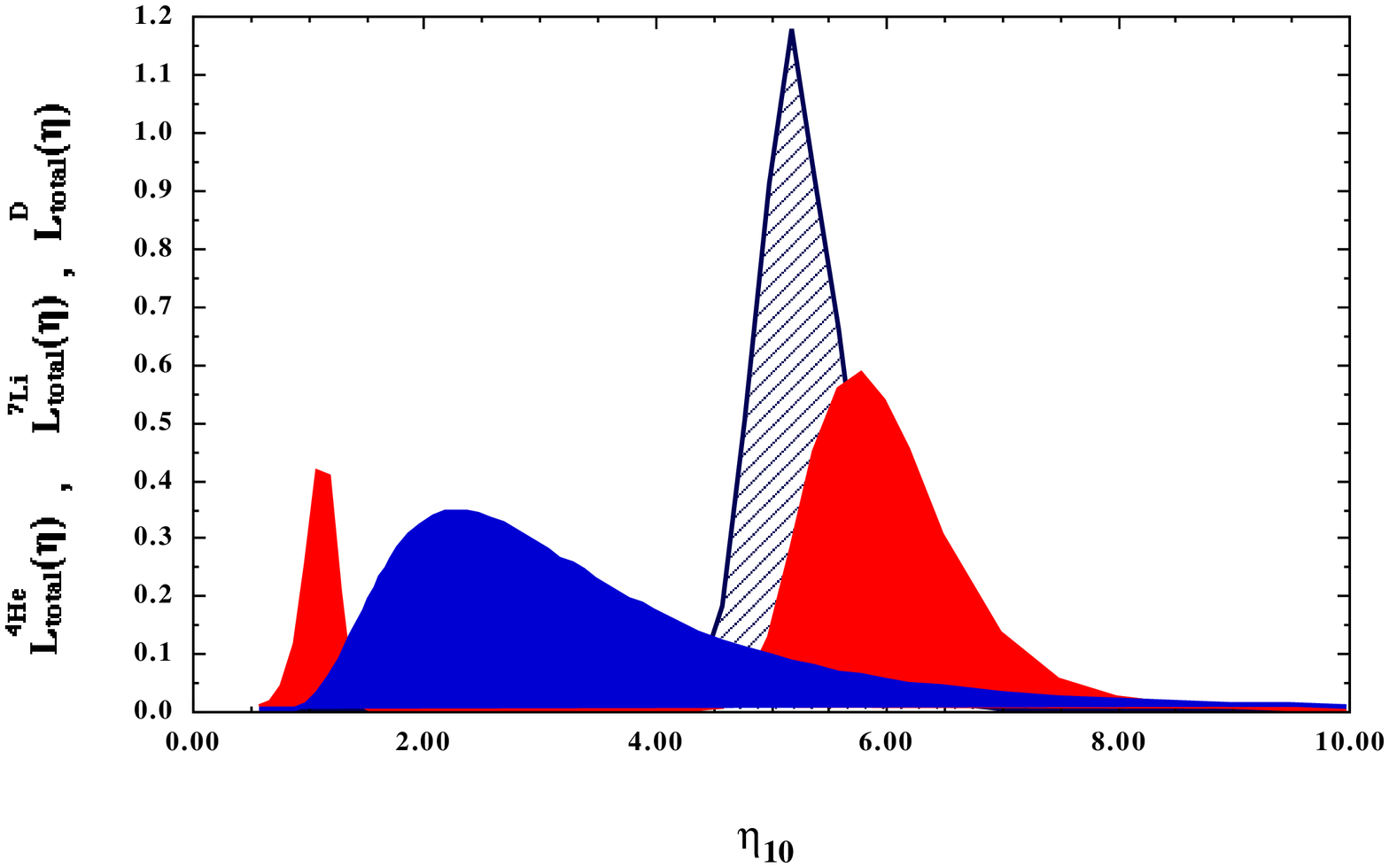}{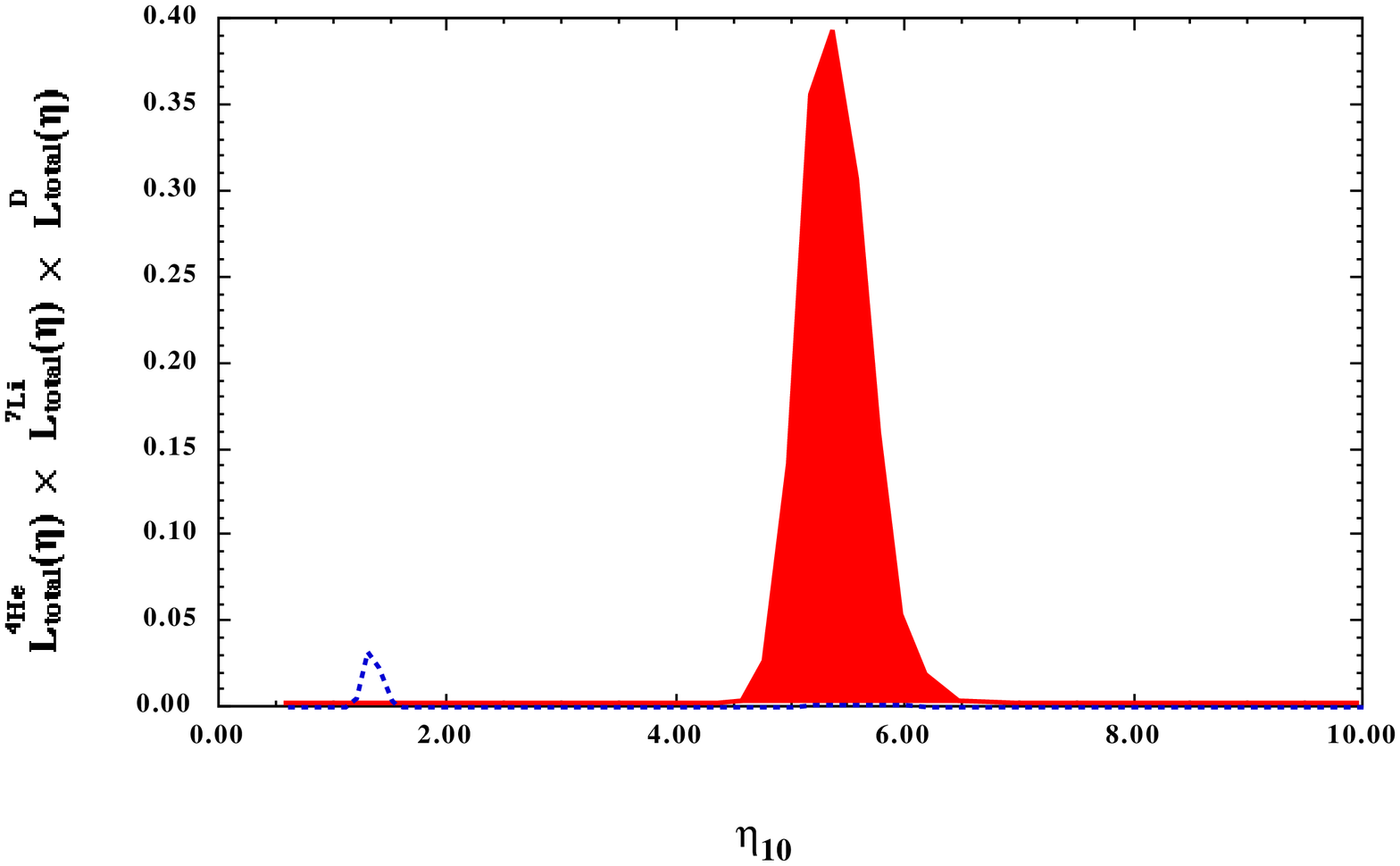} 
\caption{ (a) - As in Figure 6a, with the addition of the
likelihood  distribution for D/H assuming low D/H; (b) - The total
likelihood distribution with low D/H. The dashed curve is the distribution
from Figure 7b.}
\end{figure}

\section{Summary}

It is apparent that we have a general understanding of the production
mechanisms of the LiBeB elements.  Their (relative) inferior position on the
abundance chart indicates that they are not produced in the normal course of
stellar evolution and as is well known the mass gaps at $A = 5$ and 8 prevent
them from being produced in sufficient abundance in the big bang.
Cosmic-ray nucleosynthesis has been shown to work reasonably well in
accounting for the LiBeB abundances.  However, several questions remain.
Among them is the challenge to separate the BBN and GCRN components of \li7.
This is crucial for testing the concordance of BBN theory with observations.
The depletion of \li7 is another complication that must be resolved.
It was argued above, that \li6 may hold the key to resolving both of these
problems.  The similar processes (namely $\alpha-\alpha$ fusion) which produce
both Li isotopes could allow for a direct determination of GCRN produced \li7.
Due to its fragility, \li6, if observed in more halo stars could also play a
key role in pinning down the \li7 depletion factor. In addition, the
comparison with BeB, would be extremely useful in distinguishing between
primary and secondary models of cosmic-ray nucleosynthesis.  While great
advances have occurred in obtaining LiBeB data, open questions can only be
resolved with new data on these elements.

\acknowledgments

This work was supported in part by
DoE grant DE-FG02-94ER-40823 at the University of Minnesota.

\end{document}